\documentclass[aps,8pt,prd,twocolumn,preprintnumbers,amsmath,amssymb,nofootinbib,superscriptaddress,a4paper,showpacs]{revtex4-1}
\usepackage[colorlinks=true, pdfstartview=FitV, linkcolor=blue, citecolor=red, urlcolor=magenta]{hyperref}
\usepackage[english]{babel}%Recognize English 
\usepackage{color}%to be able to use colors
\usepackage{amsmath}
\usepackage{amssymb}
\usepackage{graphicx}
\usepackage{lmodern}
\usepackage{float}

\newcommand{\al}{\ensuremath\alpha}
\newcommand{\be}{\ensuremath\beta}
\newcommand{\ga}{\ensuremath\gamma}

\newcommand{\om}{\ensuremath\omega}

\newcommand{\beq}{\begin{equation}}
\newcommand{\eeq}{\end{equation}}
\newcommand{\bal}{\begin{align}}
\newcommand{\eal}{\end{align}}

\newcommand{\na}{\nabla}
\renewcommand{\r}{\right}
\renewcommand{\l}{\left}
\newcommand{\lr}[1]{\left(#1\right)}
\newcommand{\lrsq}[1]{\left[#1\right]}

\renewcommand{\tt}[1]{\textit{#1}}
\newcommand{\bQ}{\boldsymbol{Q}}
\newcommand{\bom}{\boldsymbol{\omega}}

\newcommand{\bg}{\boldsymbol{g}}

\newcommand{\dif}{\textbf{{d}}}

\begin{document}

\title{Conformally invariant proper time with general non-metricity}

\author{Adri\`a Delhom}
\email{adria.delhom@uv.es}
\affiliation{Departamento de F\'{i}sica Te\'{o}rica and IFIC, Centro Mixto Universidad de
Valencia - CSIC. Universidad de Valencia, Burjassot-46100, Valencia, Spain}

\author{Iarley P. Lobo}
\email{iarley\_lobo@fisica.ufpb.br}
\affiliation{Departamento de F\'isica, Universidade Federal de Lavras, Caixa Postal 3037, 37200-000 Lavras-MG, Brazil}

\author{Gonzalo J. Olmo}
\email{gonzalo.olmo@uv.es}
\affiliation{Departamento de F\'{i}sica Te\'{o}rica and IFIC, Centro Mixto Universidad de
Valencia - CSIC. Universidad de Valencia, Burjassot-46100, Valencia, Spain}
\affiliation{Departamento de F\'{\i}sica, Universidade Federal da Para\'iba\\ Caixa Postal 5008, 58051-970, Jo\~ao Pessoa, Para\'iba, Brazil}

\author{Carlos Romero}
\email{cromero@fisica.ufpb.br}
\affiliation{Departamento de F\'{\i}sica, Universidade Federal da Para\'iba\\ Caixa Postal 5008, 58051-970, Jo\~ao Pessoa, Para\'iba, Brazil}

\begin{abstract}
We show that the definition of proper time for Weyl-invariant space-times given by Perlick naturally extends to spaces with arbitrary non-metricity. We then discuss the relation between this generalized proper time and the Ehlers-Pirani-Schild definition of time when there is arbitrary non-metricity. Then we show how this generalized proper time suffers from a second clock effect. Assuming that muons are a device to measure this proper time, we constrain the non-metricity tensor on Earth's surface and then elaborate on the feasibility of such assumption.
\end{abstract}

\maketitle

%%%%%%%%%%%%%%%%%%%%%%%%%%%%%%%%%%%%%%%%%%%%%%%%%%%%%%%%%%%%%%%%%%%%%%%%%%%%%%%%%%%%%%%%%%%%%%%%%%%%%%%%%%%%%%%%%%%%%%%%%%%%%%%%%%%%%%%%%%%%%%%%%%%%%%%%%%%%%%%%%%%%%%%%%%%%%%%%%%%%%%%%%%%%%%%%%%%%%%%%%%%%%%%%

\section{Introduction}
General Relativity (GR) minimally coupled to the Standard Model has passed through different experimental tests at different scales, from precision tests in the solar system up to extreme phenomena like neutron stars or black holes mergers \cite{Abbott:2016blz,TheLIGOScientific:2017qsa}. Despite this tremendous success, there are regimes where it is necessary to assume the existence of dark components in the matter sector of the theory in order to fit the experimental data. Another possibility that could be contemplated to explain these phenomena lies in the use of alternative theories of gravity (see e.g. \cite{Heisenberg:2018vsk}). On the other hand, higher order curvature corrections to the Einstein-Hilbert action are needed to renormalize matter fields in curved spaces, and they will also be generated by quantum corrections to GR \cite{Parker:2009uva,Birrell:1982ix,Donoghue:1994dn}. One of the directions for exploring modifications of the gravitational sector comes by noticing that GR and several alternative theories have been formulated assuming a Riemannian connection. Therefore, dropping this assumption and allowing the non-Riemannian part of the affine connection to play a role in the description of space-time is a possibility that could offer interesting phenomenology to be explored. To this end, one needs to study gravitational theories in the {\it metric-affine} formulation (also called 1st order or Palatini formulation) instead of their metric formulation. The difference between both formulations is that whereas in the latter the connection is assumed to be specified by the metric (as its Levi-Civita connection), in the former the connection is an extra field that enters the action and is {\it a priori} independent of the metric. Thus the dynamics of the theory is given by extremizing the corresponding action with respect to metric and connection independently (as opposed to extremizing it only with respect to the metric field). The independence of the connection with respect to the metric introduces 64 new degrees of freedom encoded in the non-metricity $Q_{\al\mu\nu}\doteq\na_\al g_{\mu\nu}$ and the torsion $\mathcal{T}^\al{}_{\mu\nu}\doteq 2\Gamma^\al{}_{[\mu\nu]}$, which encode 40 and 24 components, respectively. By definition, these two tensors vanish in Riemannian space-times and encode the departures from metricity.\footnote{In a sense, the torsion and non-metricity tensors are a measure of how non-Riemannian a space-time is. Notice that although for some mathematicians non-Riemannian means a space-time without a Riemannian metric defined on it, for physicists the term is commonly used for space-times which have a (typically Lorenzian) metric structure, but where the connection is not compatible with the metric.}  The theoretical and phenomenological implications of including torsion and/or non-metricity in the description of space-time can be diverse, depending on the manner in which they are introduced. While the consequences of torsion have been fairly analyzed up to date, those of non-metricity have not yet been fully explored. Several classes of theories which feature torsion and/or non-metricity that have been studied in the literature are gauge theories of gravity \cite{Hehl:1994ue,Hehl:1976kj,Blagojevic:2012bc}, Ricci-Based gravity theories (which encompass Palatini $f(R)$ or Born-Infeld gravity for instance) \cite{Capozziello:2007tj,Olmo:2011uz,BeltranJimenez:2017doy,Afonso:2018bpv,Afonso:2018hyj,Afonso:2018mxn,Delhom:2019zrb,BeltranJimenez:2019acz,BeltranJimenezLargo}, teleparallel and symmetric teleparallel \cite{Hayashi:1979qx,Nester:1998mp,Obukhov:2002tm,Poltorak:2004tz,Adak:2006rx,Aldrovandi:2013wha,Maluf:2013gaa,Mol:2014ooa,BeltranJimenez:2017tkd,Combi:2017crv,BeltranJimenez:2018vdo,Koivisto:2018loq,Lobo:2019xwp,Jimenez:2019tkx,BeltranJimenez:2019tjy,Jimenez:2019ghw,Krssak:2018ywd}, hybrid gravity \cite{,Capozziello:2015lza}, Palatini scalar-tensor theories \cite{Galtsov:2018xuc}, infinite derivative theories, etc. \cite{delaCruz-Dombriz:2018aal}.\\

From a geometrical perspective, the torsion tensor measures the failure to close for infinitesimal loops built by parallel transport,  while the non-metricity tensor measures how lengths and angles are modified by parallel transport. Concretely, if we decompose the non-metricity into its irreducible components, its Weyl component $\om_\mu\doteq 1/4 Q_{\mu\al}{}^\al$ controls the change in length of a parallely transported vector. Space-times where the non-metricity is fully specified by its Weyl component are named Weyl space-times. This type of non-metricity, first introduced by Weyl \cite{weyl1}, transforms as a gauge 1-form under scale transformations of the metric, i.e. it is the gauge field associated to scale transformations (usually named dilaton field). This fact fostered the interest in Weyl geometries, since they provide a natural way of introducing scale transformations without changing the affine structure (which cannot be done in Riemannian geometries). However, although non-metricity is necessary for defining scale transformations that do not change the affine structure, the usual restriction on the non-metricity to be Weyl-like is unnecessary, and this can be achieved with general non-metricity \cite{Delhom:2019yeo}. In this case only the vectorial irreducible components of non-metricity transform as a gauge 1-form, while the tensorial irreducible components transform trivially by a conformal factor\footnote{Concretely under a conformal transformation  $\bg\mapsto\tilde\bg=e^\phi\bg$ in 4 space-time dimensions, the different irreducible components listed in \eqref{decompNM} transform as: $\tilde Q_{1\mu}=Q_{1\mu}+4 (d\phi)_\mu$ , $\tilde Q_{2\mu}=Q_{1\mu}+ (d\phi)_\mu$, $\tilde S_{\mu\al\be}=e^\phi S_{\mu\al\be}$ and $M_{\mu\al\be}=e^\phi M_{\mu\al\be}$}.\\

 Motivated by the discussion of Perlick  \cite{PerlickTime} about how to construct a suitable Weyl-invariant notion of proper time in Weyl-invariant space-times that reduces to the standard definition of proper time in the Riemannian limit, and following  \cite{Lobo:2018zrz} in its analysis of the physical role played by the Weyl 1-form, we will be concerned about finding a suitable definition of proper time that respects scale invariance in the presence of arbitrary non-metricity (or generalized Weyl invariance in the sense of \cite{Delhom:2019yeo}), and also with the physical consequences of having non-trivial non-metricity if physical time was described by this definition. To that end we will generalize the parametrization for generalized proper time found in \cite{Avalos:2016unj} to the case of arbitrary non-metricity and find the existence of a conformally invariant second clock effect related to an arbitrary non-metricity tensor. Then, under the assumption that fundamental particles measure the generalized proper time,  we show that it is possible to constrain some components of the non-metricity around Earth's surface by considering data on the time dilation of muons accelerated by a magnetic field, which give constraints to the amount of second clock effect that we would find by considering different muon trajectories in a background with non-metricity. We also discuss these bounds in several theories featuring non-metricity. Finally we elaborate on the conditions that should be satisfied by the matter fields and their coupling to the geometry so that generalized clocks can be built by means of the M\"arkze-Wheeler construction \cite{MarkzeWheeler,Desloge1989-DESATD}, which would make generalized proper time observable.\\
 
Our paper is organized as follows. In section (\ref{sec:GenTime}) we  derive the formula that gives the generalized proper time interval between two events connected by a given curve. In section (\ref{sec:EPS}) we discuss the relation between generalized proper time and the operational definition of proper time given by Ehlers, Pirani and Schild (EPS) in \cite{EPS}. In section (\ref{sec:SecondClock}) we show how from the definition of generalized proper time one can deduce that in the presence of non-metricity there will be an effect in the time measured by an ideal generalized clock that depends on the whole history of the clock. This effect is usually named {\it second clock effect}. We also point out what irreducible components of the non-metricity tensor would contribute to this second clock effect and which would not. In section (\ref{sec:Muon}) we analyze the physical consequences that would occur if muons were generalized clocks by using data on experiments that measure the dilation of the lifetime of the muon when accelerated in a constant magnetic field. For this purpose we assume a constant background non-metricity around Earth's surface and then particularize to specific theories. Finally, in section (\ref{sec:Conc}) we discuss under which conditions generalized proper time could be physical, in the sense that of whether it is possible to build clocks that measure it and some other final remarks.

%%%%%%%%%%%%%%%%%%%%%%%%%%%%%%%%%%%%%%%%%%%%%%%%%%%%%%%%%%%%%%%%%%%%%%%%%%%%%%%%%%%%%%%%%%%%%%%%%%%%%%%%%%%%%%%%%%%%%%%%%%%%%%%%%%%%%%%%%%%%%%%%%%%%%%%%%%%%%%%%%%%%%%%%%%%%%%%%%%%%%%%%%%%%%%%%%%%%%%%%%%%%%%%%

%%%%%%%%%%%%%%%%%%%%%%%%%%%%%%%%%%%%%%%%%%%%%%%%%%%%%%%%%%%%%%%%%%%%%%%%%%%%%%%%%%%%%%%%%%%%%%%%%%%%%%%%%%%%%%%%%%%%%%%%%%%%%%%%%%%%%%%%%%%%%%%%%%%%%%%%%%%%%%%%%%%%%%%%%%%%%%%%%%%%%%%%%%%%%%%%%%%%%%%%%%%%%%%%

\section{Generalized proper time}\label{sec:GenTime}

The usual Riemannian proper time, which is defined as the arc-length of time-like curves, is not invariant under scale (or Weyl) transformations. Worried with the possibility of studying the physics that occurs in Weyl spaces,\footnote{We will use Weyl space-time meaning Weyl-invariant space-time from here on.} Perlick coined a way to define a proper time in Weyl spaces that is Weyl-invariant and reduces to the standard Riemannian proper time in the appropriate limit in a canonical way, {\it i.e.} using only the metric and the connection that define the Weyl space  \cite{PerlickTime}. We will call this a {\it generalized proper time}. More recently, it was shown \cite{Avalos:2016unj} that generalized proper time coincides with the operational time given by EPS in \cite{EPS}, where they deduced from an operational point of view, and under certain assumptions, that the space-time manifold could be described by a Weyl space. In \cite{Avalos:2016unj} a formula for computing the proper time between two causally connected events was also given as follows.\\

The proper time interval between two events $\ga(t_0)$ and $\ga(t)$ belonging to a time-like curve $\ga: t\in I \mapsto \ga(t)\in M$ is given by
\begin{align}\label{PerlickTimeEq}
& \Delta\tau(t)=\\
& \frac{\l.\frac{d\tau}{dt}\r|_{(t=t_0)}}{\sqrt{g\lr{\ga'(t_0),\ga'(t_0)}}}\int_{t_0}^t \sqrt{g\lr{\ga'(u),\ga'(u)}}e^{-\frac{1}{2}\int_{u_0}^u \om\lr{\ga'(s)}ds}du \;,\nonumber
\end{align}
where $\ga'=d\ga (t)/dt$, and in a Weyl space the compatibility condition between metric and affine structures is given by $\na{\boldsymbol{g}}\doteq{\boldsymbol{Q}}=\boldsymbol{\om}\otimes \boldsymbol{g}$ (where $\nabla$ is the covariant derivative and $\otimes$ is the usual tensor product). We are here concerned with finding a definition of proper time in scale-invariant space-times with a general form of the non-metricity tensor $\bQ$, {\it i.e.} in generalized Weyl space-times in the sense of \cite{Delhom:2019yeo}. In order to do so, we will start with the same definition as presented by Perlick in \cite{PerlickTime}, which can be generalized to an arbitrarily general space-time in a straightforward way:\\

\tt{A $\tau$-parametrized time-like curve $\ga: \tau\in I \mapsto \ga(\tau)\in M$ is a generalized clock if}
\beq\label{DefGenTime}
g\lr{\ga'(\tau),\frac{D\ga'}{d\tau}}=0 \quad \forall\tau\in I.
\eeq
\tt{\quad{}The parameter $\tau$ parametrizing a generalized clock is the generalized proper time measured by the clock.}\\

In this definition, for any vector field $V$, the term $DV/dt$ is the covariant derivative of $V$ along $\gamma'$, i.e., $DV/dt\doteq \nabla_{\ga'}V$, where $\nabla_{\spadesuit}\clubsuit$ is the covariant derivative of $\clubsuit$ in the direction of $\spadesuit$.
\par
It can be shown that every time-like path $\ga$ admits a parametrization with generalized proper time, {\it i.e} every physical observer can be a generalized clock \cite{Avalos:2016unj}. In their argument the authors start with a time-like curve $\ga(t)$ and show that a reparametrization $\mu : t \in I \mapsto \mu(t)\doteq\tau\in I'$ satisfying

\beq\label{ConditionRepClock}
\frac{d^2\mu}{dt^2}-\frac{g\lr{\ga'(t),\frac{D\ga'(t)}{dt}}}{g\lr{\ga'(t),\ga'(t)}}\frac{d\mu}{dt}=0
\eeq
has the property that $\tilde{\ga}(\tau)=\ga\circ\mu^{-1}(\tau)$ is a generalized clock. As \eqref{ConditionRepClock} always has a unique solution, every observer (time-like curve) can be a generalized clock. The proof outlined in \cite{Avalos:2016unj} is independent of the relation between metric and affine structure, which allows us to use this result and follow the steps of \cite{Avalos:2016unj} to find a general solution for \eqref{ConditionRepClock} in a space-time with arbitrary $\boldsymbol{Q}$. In order to derive this solution, let us start from  the definition of the covariant derivative for rank 2 tensor fields
\beq\label{CompatibiltyCondition}
\lr{\na_X g}(Y,Z)= X\lr{g\lr{Y,Z}}-g\lr{\na_X Y,Z}-g\lr{Y,\na_XZ},
\eeq
where $X,\, Y,\, Z$ are three arbitrary vector fields. Notice that by definition of the non-metricity tensor, $Q\lr{X,Y,Z}\doteq\lr{\na_X g}(Y,Z)$. Since the relation $\na_{\ga'}=d/dt$ is satisfied along the curve $\ga(t)$, using  \eqref{CompatibiltyCondition} with $X=Y=Z=\ga'(t)$ leads to
\beq
\frac{g\lr{\ga',\frac{D\ga'}{dt}}}{g\lr{\ga',\ga'}}=\frac{1}{2}\lrsq{\frac{d}{dt} \ln\lr{g\lr{\ga',\ga'}}-\frac{Q\lr{\ga',\ga',\ga'}}{g\lr{\ga',\ga'}}}\:,
\eeq
which is analogous to Eq.(9) of \cite{Avalos:2016unj} after the substitution\footnote{Notice that this substitution is a particularization of a general non-metricity tensor to a Weyl-type non-metricity tensor.} $Q(\ga',\ga',\ga')\mapsto \om\lr{\ga'}g\lr{\ga',\ga'}$. Which combined with equation (\ref{ConditionRepClock}) gives \\
\beq\label{cond_add}
\frac{d\mu}{dt}=\frac{d\mu(t_0)}{dt}\left[\frac{g(\gamma'(t),\gamma'(t))}{g(\gamma'(t_0),\gamma'(t_0))}\right]^{1/2}e^{-\frac{1}{2}\int_{t_0}^t \frac{Q(\gamma'(s),\gamma'(s),\gamma'(s))}{g(\gamma'(s),\gamma'(s))}ds}.
\eeq

Integrating this equation for $\mu=\tau$ leads to a formula for computing the generalized proper time as defined in \eqref{DefGenTime} elapsed between two events $A=\ga(t_0)$ and $B=\ga(t)$ for the observer $\ga(t)$
\begin{widetext}
\beq\label{GeneralProperTimeEq}
\Delta\tau(t)=\frac{\l.\frac{d\tau}{dt}\r|_{t=t_0}}{\sqrt{g\lr{\ga'(t_0),\ga'(t_0)}}}\int_{t_0}^t\sqrt{g\lr{\ga'(u),\ga'(u)}}e^{-\frac{1}{2}\int_{u_0}^u\frac{Q\lr{\ga'(s),\ga'(s),\ga'(s)}}{g\lr{\ga'(s),\ga'(s)}}ds}du.
\eeq
\end{widetext}

This formula reduces to \eqref{PerlickTimeEq} if the non-metricity tensor is specified to be of the Weyl kind,  which is the corresponding one for Perlick time\footnote{Note that Perlick time is just generalized proper time with the restriction that the non-metricity is Weyl-like.} as found in \cite{Avalos:2016unj}. The proof of the additivity of generalized proper time as specified by \eqref{GeneralProperTimeEq} is completely analogous to that outlined in \cite{Avalos:2016unj}. Due to the fact that \eqref{GeneralProperTimeEq} reduces to \eqref{PerlickTimeEq} for Weyl non-metricity, and as proven in \cite{Avalos:2016unj} \eqref{PerlickTimeEq} has the correct Weyl-Integrable-space-time (WIST) and Riemannian limits, this general proper time \eqref{GeneralProperTimeEq} will also have the correct WIST and Riemannian limits, thus being a sensible generalization of Riemannian proper time.
\par
In Ref.\cite{Delhom:2019yeo}, it was found that in non-Riemannian manifolds, under a scale transformations of the metric, the scale invariance of the affine structure implies certain transformation properties of the non-metricity tensor under scale transformations. In fact, one can verify that the simultaneous transformations
\begin{equation}\label{conf-trans}
\tilde{\boldsymbol{g}}=e^{\phi}\boldsymbol{g} \ \ \ \ \text{and} \ \ \ \ \tilde{\boldsymbol{Q}}=e^{\phi}(\boldsymbol{Q}+\boldsymbol{d}\phi\otimes \boldsymbol{g})
\end{equation}
leave invariant the affine connection (as scale transformations should), where $\phi$ is any arbitrary smooth scalar function. Since conformal transformations do not modify the orthogonality conditions (\ref{DefGenTime}) is automatically preserved by these transformations and therefore the definition of generalized proper time naturally encodes conformal invariance, which can also be verified by using \eqref{conf-trans} on the expression \eqref{GeneralProperTimeEq} for generalized proper time. Thus, the aforementioned properties of \eqref{GeneralProperTimeEq} suggest that it is a suitable generalization of proper time for conformally invariant space-times with general non-metricity.

%%%%%%%%%%%%%%%%%%%%%%%%%%%%%%%%%%%%%%%%%%%%%%%%%%%%%%%%%%%%%%%%%%%%%%%%%%%%%%%%%%%%%%%%%%%%%%%%%%%%%%%%%%%%%%%%%%%%%%%%%%%%%%%%%%%%%%%%%%%%%%%%%%%%%%%%%%%%%%%%%%%%%%%%%%%%%%%%%%%%%%%%%%%%%%%%%%%%%%%%%%%%%%%%

\section{Relation between generalized proper time and EPS proper time}\label{sec:EPS}

In the framework introduced by EPS in \cite{EPS} one of the key assumptions that lead to the conclusion that space-time should be a Weyl space\footnote{Of course, Riemannian space is a sub case of Weyl space with $\bom=0$} was the compatibility between the projective structure given by freely falling particles and the conformal structure given by the light rays. They also define a notion of proper time within this framework which is Weyl invariant and coincides with generalized proper time in Weyl spaces \cite{Avalos:2016unj}. Thus, under the restriction to the non-metricity tensor to be Weyl-like, the generalized proper time should boil down to EPS proper time. Let us find out whether the equivalence between EPS and generalized proper times can also be achieved for more general kinds of non-metricity, or rather Weyl-like non-metricity is the most general form of non-metricity that allows this.  To that end, we proceed by generalizing the proof given in \cite{Avalos:2016unj} for the equivalence of EPS and Perlick clocks.\\

Let us first study under which conditions an EPS clock is also a generalized clock. By definition, a time-like curve $\ga(\tau)$ is an EPS clock ({\it i.e.} it is parametrized by EPS time) if there exists a vector field $V_\ga(\tau)$ which is parallel along $\ga(\tau)$ and satisfies $g(\ga'(\tau),\ga'(\tau))=g(V_\ga(\tau),V_\ga(\tau))$ along the curve \cite{EPS}. Differentiating this condition one finds from \eqref{CompatibiltyCondition} and using the fact that $V_\ga(\tau)$ is parallely transported along $\ga(\tau)$, so that  $DV_\ga(\tau)/d\tau=0$, the following relation follows

\begin{align}\label{ConditionEPSisPerlick}
&2g\lr{\frac{D\ga'(\tau)}{d\tau},\ga'(\tau)}=\\
&Q(\ga'(\tau),V_\ga(\tau),V_\ga(\tau))-Q(\ga'(\tau),\ga'(\tau),\ga'(\tau)).\nonumber
\end{align}

This condition is valid for any EPS clock. Then for an EPS clock to be also a generalized clock, by definition of generalized clock, the condition $Q(\ga'(\tau),\ga'(\tau),\ga'(\tau))=Q(\ga'(\tau),V_\ga(\tau),V_\ga(\tau))$ must hold for all time-like curves where $V_\ga(\tau)$ is the vector field that satisfies $g(\ga'(\tau),\ga'(\tau))=g(V_\ga(\tau),V_\ga(\tau))$ along each timelike curve $\ga(\tau)$. Let us now see the conditions needed for a generalized clock to be an EPS clock. By definition, a time-like curve $\ga(\tau)$ is a generalized clock if $\ga'(\tau)$ and $D\ga'(\tau)/d\tau$ are orthogonal along the curve. Define (locally) a parallel vector field $V_\ga(\tau)$ along $\ga(\tau)$ as the unique solution to the the initial value problem

\beq\label{InitialValueProblem}
\frac{DV_\ga(\tau)}{d\tau}=0\ \ \ \text{with initial condition}\ \ \ V_\ga(\tau_0)=\ga'(\tau_0).
\eeq

Using that such a $V_\ga(\tau)$ is parallely transported along $\ga(\tau)$ and the orthogonality between $\ga'(\tau)$ and $D\ga'(\tau)/d\tau$, from \eqref{CompatibiltyCondition} one finds

\begin{align}
&\frac{d}{d\tau}g\lr{V_\ga(\tau),V_\ga(\tau)}=Q(\ga'(\tau),V_\ga(\tau),V_\ga(\tau)),\\
&\frac{d}{d\tau}g\lr{\ga'(\tau),\ga'(\tau)}=Q(\ga'(\tau),\ga'(\tau),\ga'(\tau)),
\end{align}
which together with the initial condition $g\lr{V_\ga(\tau_0),V_\ga(\tau_0)}=g\lr{\ga'(\tau_0),\ga'(\tau_0)}$  in \eqref{InitialValueProblem} define a unique solution for $g\lr{V_\ga(\tau),V_\ga(\tau)}$ and $g\lr{\ga'(\tau),\ga'(\tau)}$ respectively. These solutions will satisfy $g\lr{V_\ga(\tau),V_\ga(\tau)}=g\lr{\ga'(\tau),\ga'(\tau)}$ only if the condition

\beq\label{ConditionPerlickisEPS}
Q(\ga'(\tau),V_\ga(\tau),V_\ga(\tau))=Q(\ga'(\tau),\ga'(\tau),\ga'(\tau))
\eeq
is also satisfied along $\ga(\tau)$. Then for a generalized clock to be also an EPS clock \eqref{ConditionPerlickisEPS} must hold for all time-like curves. And therefore the condition \eqref{ConditionPerlickisEPS} is a necessary and sufficient condition for a generalized clock to be an EPS clock. We can now try to figure out the solutions for a non-metricity tensor that satisfies the above condition. First we re-write this condition in the following way: For every timelike curve (remember every {\it timelike} path can be a Perlick clock) define a symmetric  2-tensor by $q^{(\ga)}{}_{\al\be}(\tau)\doteq Q_{\mu\al\be}\ga'^\mu(\tau)$. Thus condition \eqref{ConditionPerlickisEPS} can be restated as 
\begin{equation}\label{CondQEPS=Perlick}
q^{(\ga)}{}_{\al\be}\lr{V_{(\ga)}{}^\al V_{(\ga)}{}^\be-\ga'^\al\ga'^\be}=0,
\end{equation}
where $V_{(\gamma)}{}^{\alpha}$ denote the components of the $V_{\gamma}$. This condition is an algebraic constraint that should be satisfied by $q^{(\ga)}{}_{\al\be}$ for every EPS and generalized clock $\ga(\tau)$, and if it is satisfied every EPS clock would also be a generalized clock and vice-versa. Now notice that for a given clock $\ga(\tau)$, and at a given time, there is only one constraint given by \eqref{ConditionPerlickisEPS} and 10 (algebraicaly) independent components of $q^{(\ga)}{}_{\al\be}$. Thus the system is indeterminate. Since it is homogeneous, a Riemannian space-time ($\boldsymbol{q}^{(\ga)}=0 \;\forall \, \ga(\tau), \tau \Rightarrow \bQ=0$) is a solution. However, there is another canonical solution that can be found by using one of the conditions for $\ga(\tau)$ to be an EPS clock, which can be written as $g_{\al\be}\lr{V_{(\ga)}{}^\al V_{(\ga)}{}^\be-\ga'^\al\ga'^\be}=0$. From this condition, it is apparent that $\boldsymbol{q}^{(\ga)}=\al_\ga \bg|_{\ga(\tau)}$ where $\al_\ga$ is a $\ga(\tau)$-dependent proportionality factor is also a solution,\footnote{Here $|_{\ga(\tau)}$ means evaluated at the point $\ga(\tau)$.} which can also be written $Q_{\mu\al\be}\ga'^\mu(\be)=\al_\ga g_{\al\be}$. In this case, we can always find at each point a 1-form $\bom$ such that $\al_\ga=\omega(\ga'(\tau))$ for all the curves through that point. Thus this other canonical solution is nothing but a Weyl space-time with $\bQ=\bom\otimes\bg$. One further way of understanding whether there can be other solutions is to look at the conditions on the irreducible components of $\bQ$. In 4 space-time dimensions, the non-metricity tensor can be decomposed in its Lorentz-irreducible pieces as
\begin{align}\label{NMdecomposed}
&Q_{\mu\alpha\beta}=\frac{1}{18}(5Q_{1\mu} g_{\alpha\beta}-Q_{1\alpha} g_{\beta\mu}-Q_{1\beta} g_{\mu\alpha}-2Q_{2\mu} g_{\alpha\beta}\nonumber\\
&+4Q_{2\alpha} g_{\beta\mu}+4Q_{2\beta} g_{\mu\alpha})+S_{\mu\alpha\beta}+M_{\mu\alpha\beta},
\end{align}
where
\begin{align}
\begin{split}\label{decompNM}
&Q_{1\mu}\doteq  g^{\alpha\beta}Q_{\mu\alpha\beta} \ \ \  , \ \ \ Q_{2\mu}\doteq  g^{\alpha\beta}Q_{\alpha\mu\beta},\\
&S_{\mu\alpha\beta}\doteq \frac{1}{3}(Q_{\mu\alpha\beta}+Q_{\alpha\beta\mu}+Q_{\beta\mu\alpha})-\frac{1}{18}(Q_{1\mu} g_{\alpha\beta}\\
&+Q_{1\alpha} g_{\beta\mu}+Q_{1\beta} g_{\mu\alpha})-\frac{1}{9}(Q_{2\mu} g_{\alpha\beta}+Q_{2\alpha} g_{\beta\mu}+Q_{2\beta} g_{\mu\alpha}),\\
&M_{\mu\alpha\beta}\doteq \frac{1}{3}(2Q_{\mu\alpha\beta}-Q_{\alpha\beta\mu}-Q_{\beta\mu\alpha})-\frac{1}{9}(2Q_{1\mu} g_{\alpha\beta}\\
&-Q_{1\alpha} g_{\beta\mu}-Q_{1\beta} g_{\mu\alpha})+\frac{1}{9}(2Q_{2\mu} g_{\alpha\beta}-Q_{2\alpha} g_{\beta\mu}-Q_{2\beta} g_{\mu\alpha}).
\end{split}
\end{align}

After some algebra, we can see that \eqref{ConditionPerlickisEPS} leads to a relation between vectorial and tensorial components of non-metricity that reads
\begin{equation}\label{condEPSPerlick2}
\ga'^\mu\lr{S_{\mu\al\be}+M_{\mu\al\be}}=\frac{1}{9}\ga'_{(\al}\lr{4Q_{2\be)}-Q_{1\be)}}.
\end{equation}
Notice that this relation has to be satisfied for every timelike $\ga'(\tau)$, and given a point, there are infinite $\ga'(\tau)$ at that point, but the non-metricity components have to satisfy \eqref{condEPSPerlick2} for all such tangent vectors. Thus, because of the tensor structure of the equation, the most general way to satisfy \eqref{condEPSPerlick2} for all tangent vectors is to have $\boldsymbol{M}=-\boldsymbol{S}$ and $\boldsymbol{Q}_1=4\boldsymbol{Q}_2$. Using these conditions into \eqref{NMdecomposed}, we are led to a form of the non-metricity tensor written in terms of its irreducible components as
\begin{equation}
\bQ=\bQ_2\otimes\bg=\frac{1}{4}\bQ_1\otimes\bg,
\end{equation}
which implies that the most general space-time where any EPS clock is a generalized clock and vice-versa is a Weyl space-time. In fact, the EPS paper states that from its construction based on the compatibility of the projective and conformal structure one is led naturally to a Weyl geometry \cite{EPS} (although some subtleties have been  addressed in \cite{Trautman2012,Matveev:2013fpl,Scholz:2019tif,Matveev:2020wif}). That is the reason why the EPS time is irrevocably connected to the geometrical definition of time that uses the metric and the connection introduced by Perlick, and if one wants to study a natural notion of time in a space-time where free particles follow geodesics of a general non-metric connection, generalized proper time can be a good candidate from the theoretical point of view, since it suitably generalizes Riemannian and Perlick times. In the next sections we are going to present an important physical effect resulting from the use of this geometrical time: the second clock effect. That was already present in the original version of Weyl's theory and, in fact, persists in the general non-metricity case.

%%%%%%%%%%%%%%%%%%%%%%%%%%%%%%%%%%%%%%%%%%%%%%%%%%%%%%%%%%%%%%%%%%%%%%%%%%%%%%%%%%%%%%%%%%%%%%%%%%%%%%%%%%%%%%%%%%%%%%%%%%%%%%%%%%%%%%%%%%%%%%%%%%%%%%%%%%%%%%%%%%%%%%%%%%%%%%%%%%%%%%%%%%%%%%%%%%%%%%%%%%%%%%%%

\section{Generalized proper time and the second clock effect}\label{sec:SecondClock}

\begin{figure}[H]
\hfill\includegraphics[scale=0.25]{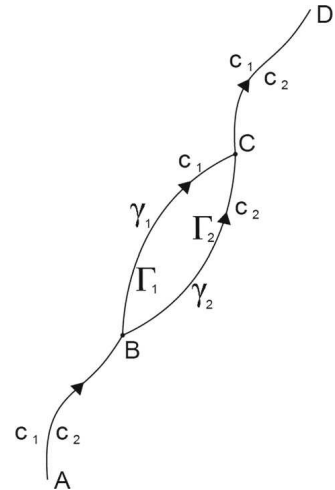}\hspace*{\fill}%
\caption{Synchronized clocks $c_{1}$ and $c_{2}$ follow world lines
$\gamma_{1}$ and $\gamma_{2}$, which are coincident from point $A$ to $B$,
where they are separated to follow the parts $\Gamma_{1}$ and $\Gamma_{2}$ of
these lines until point $C$, where they are once again joined and continue
together until point $D$.}%
\label{fig1}%
\end{figure}

As a postscript to the original paper by Weyl \cite{weyl1}, in which he introduces his theory with a geometrization of electromagnetism by means of a Weyl-like non-metricity tensor, Einstein criticized the proposal by stating that the theory had an unpleasant effect due to the non-integrability of lengths. The clock rate of standard clocks would depend on their past histories, which would have imprints on the spectral lines of atoms (an effect that was not observed). This effect was later coined as the {\it second clock effect} \cite{brown}. In order to illustrate it, consider two clocks $c_{1}$ and $c_{2}$ synchronized at point $A$ (see Fig.(\ref{fig1})), which are transported together until point $B$, then separated and transported along two different paths, $\Gamma_{1}$ and $\Gamma_{2}$, until point $C$, where they are rejoined.

Let the proper time starting from point $C$ to be 
\begin{widetext}
\begin{align}
\tau =\frac{\tau'(u_C)}{(g(\gamma'(u_C),\gamma'(u_C)))^{1/2}}\int_{u_C}^u \sqrt{g(\gamma'(u),\gamma'(u))} e^{-\frac{1}{2}\int_{u_C}^u\frac{Q(\gamma'(s),\gamma'(s),\gamma'(s))}{g(\gamma'(s),\gamma'(s))}ds},\label{time1}\\
\bar{\tau}=\frac{\bar{\tau}'(u_C)}{(g(\gamma'(u_C),\gamma'(u_C)))^{1/2}}\int_{u_C}^u \sqrt{g(\gamma'(u),\gamma'(u))} e^{-\frac{1}{2}\int_{u_C}^u\frac{Q(\gamma'(s),\gamma'(s),\gamma'(s))}{g(\gamma'(s),\gamma'(s))}ds},\label{time2}
\end{align}
\end{widetext}
where $\tau$ and $\bar{\tau}$ are the measuring of clocks $1$ and $2$, respectively. 
\par
The times from $A$ to $C$ can be computed using Eq.(\ref{cond_add}) as
\begin{align}
&\tau'(u_C)=\label{cond1}\\
&\tau'(u_A)\left[\frac{g(\gamma'_1(u_C),\gamma'_1(u_C))}{g(\gamma'_1(u_A),\gamma'_1(u_A))}\right]^{\frac{1}{2}}e^{-\frac{1}{2}\int_{u_A}^{u_C} \frac{Q(\gamma'_1(s),\gamma'_1(s),\gamma'_1(s))}{g(\gamma'_1(s),\gamma'_1(s))}ds},\nonumber\\
&\bar{\tau}'(\bar{u}_C)=\label{cond2}\\
&\bar{\tau}'(\bar{u}_A)\left[\frac{g(\gamma'_2(\bar{u}_C),\gamma'_2(\bar{u}_C))}{g(\gamma'_2(\bar{u}_A),\gamma'_2(\bar{u}_A))}\right]^{\frac{1}{2}}e^{-\frac{1}{2}\int_{\bar{u}_A}^{\bar{u}_C} \frac{Q(\gamma'_2(s),\gamma'_2(s),\gamma'_2(s))}{g(\gamma'_2(s),\gamma'_2(s))}ds},\nonumber
\end{align}
where $(u,\gamma_1)$ and $(\bar{u},\gamma_2)$ are the parameters and world lines of paths $1$ and $2$. As in \citep{Avalos:2016unj}, after a reparametrization from $\bar{u}$ to $u$, we find
\begin{align}\label{cond3}
&\bar{\tau}'(u_C)=\\
&\bar{\tau}'(u_A)\left[\frac{g(\gamma'_2(u_C),\gamma'_2(u_C))}{g(\gamma'_2(u_A),\gamma'_2(u_A))}\right]^{\frac{1}{2}}e^{-\frac{1}{2}\int_{\bar{u}_A}^{\bar{u}_C} \frac{Q(\gamma'_2(s),\gamma'_2(s),\gamma'_2(s))}{g(\gamma'_2(s),\gamma'_2(s))}ds}.\nonumber
\end{align}
Then, using the fact that $\ga'_1(u_A)=\ga'_2(u_A)$, $\ga'_1(u_C)=\ga'_2(u_C)=\ga'(u_C)$ and placing Eqs.(\ref{cond1}) and (\ref{cond3}) into Eq.(\ref{time2}), we find
\begin{align}
&\bar{\tau}=\\
&\tau\frac{\bar{\tau}'(u_A)}{\tau'(u_A)}e^{\frac{1}{2}\int_{u_A}^{u_C} \frac{Q(\gamma'_1(s),\gamma'_1(s),\gamma'_1(s))}{g(\gamma'_1(s),\gamma'_1(s))}ds-\frac{1}{2}\int_{\bar{u}_A}^{\bar{u}_C} \frac{Q(\gamma'_2(s),\gamma'_2(s),\gamma'_2(s))}{g(\gamma'_2(s),\gamma'_2(s))}ds}.\nonumber
\end{align}

Since both clocks have the same scale at the event $A$, i.e., $\bar{\tau}'(u_A)=\tau'(u_A)$. Therefore, a clock that measures generalized proper time will measure a second clock effect for a general non-metricity tensor given by
\beq\label{SecondClockEffect}
\bar{\tau}=\tau \exp\left(\frac{1}{2}\int_{\Gamma_1-\Gamma_2}\frac{Q(\gamma'(s),\gamma'(s),\gamma'(s))}{g(\gamma'(s),\gamma'(s))}ds\right).
\eeq
Notice that this expression reduces to the result found in \cite{Avalos:2016unj} for a Weyl-like non-metricity tensor $\boldsymbol{Q}=\boldsymbol{\om}\otimes \boldsymbol{g}$. Also notice that such relation is invariant under the action of conformal transformations \eqref{conf-trans}, $\int_{\Gamma_1-\Gamma_2} d\phi=0$ ( $\Gamma_1-\Gamma_2$ is a closed path). Therefore our construction describes a conformally invariant second clock effect. %%%%%%%%%%%%%%%%%%%%%%%%%%%%%%%%%%%%%%%%%%%%%%%%%%%%%%%%%%%%%%%%%%%%%%%%%%%%%%%%%%%%%%%%%%%%%%%%%%%%%%%%%%%%%%%%%%%%%%%%%%%%%%%%%%%%%%%%%%%%%%%%%%%%%%%%%%%%%%%%%%%%%%%%%%%%%%%%%%%%%%%%%%%%%%%%%%%%%%%%%%%%%%%%

\section{Observability of the second clock effect}\label{sec:Muon}

\begin{figure}[H]
\hfill\includegraphics[scale=0.3]{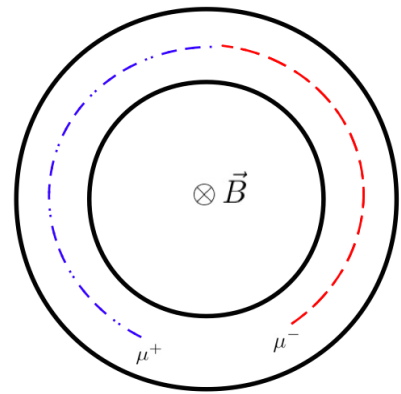}\hspace*{\fill}\caption{For a
magnetic field $\vec{B}$ into the page, the positive muon $\mu^{+}$ follows
the circle in blue (dot-dashed line), while the negative muon follows the red
circle (dashed line). }%
\label{fig2}%
\end{figure}

We saw in the last section that generalized clocks suffer from the second clock effect. This way, experiments that test the usual proper time formula are natural candidates for testing the existence of this effect. Since the second clock effect is complementary to the first clock effect (time dilation), experiments designed to test the latter should present imprints of the former. For instance, the experiments that manifest the dilation of the lifetime of fundamental particles constitute a natural environment for our investigations. In this direction, the most suitable experiments of us are those that test the anomalous magnetic moment of the muon, which were already used in \cite{Lobo:2018zrz} by assuming that the Weyl 1-form was proportional to the electromagnetic four-potential. These experiments are usually made by accelerating a beam of particles encapsulated in a Muon Storage Ring, where their dilated lifetimes are measured with high precision. Let us review how the first clock effect manifests in this context. Consider a muon accelerated by a constant magnetic field $\vec{B}$ in the $z$-direction. Due to the Lorentz force, the particle follows a circular trajectory with constant velocity (Fig. \ref{fig2}). Depending on the particle's electric charge it will follow a clockwise or a counter-clockwise direction with $\vec{v}=\pm v_0\, \hat{\varphi}$ (we are considering cylindrical coordinates, for simplicity).

If we consider Minkowski space-time (with vanishing non-metricity), the proper time of the muon is 
\beq
\tau(u)=c^{-1}\int_{u_0}^{u}\sqrt{g(\gamma',\gamma')}\, ds.
\eeq

If we use the coordinate $x^0=c\, t$ as parameter and setting the initial time as $t_0=0$, we have
\beq
\tau(t)=\int_{0}^t\sqrt{1-v_0^2/c^2}\, dt'=\int_0^t\gamma^{-1}\, dt',
\eeq
where we used the Lorentz factor $\gamma^{-1}=\sqrt{1-v_0^2/c^2}$. Do not confuse with the $\gamma$-symbol that we are using for space-time curves in this paper. Notice that the coordinate time, $t$, is the time measured by the observer in the laboratory frame. For a circular trajectory with constant velocity, the lifetime of the muon (as measured by an observer in the laboratory) is thus dilated by the Lorentz factor as
\beq
\Delta t=\gamma \Delta \tau.
\eeq

For experiments that measure the anomalous magnetic moment of the muon we usually have $\gamma\approx 29.3$, which is called the ``magic $\gamma$'' due to the fact that, for this value of the Lorentz factor, the contribution of a focusing electrostatic potential is removed from the rotation of the muon magnetic moment \cite{jegerlehner2008anomalous}.
\par
In the Muon Storage Ring at CERN, the anomalous magnetic moment of the muon was measured \cite{Bailey:1978mn}, and it was also measured the relativistic dilation of its lifetime \cite{Bailey:1977de}. At rest, the muon has lifetime $\tau \approx 2.2\, \mu s$, and in this experiment it is dilated to $t\approx 64.4\, \mu s$. This experimental result is also consistent with the CPT symmetry that predicts that $\mu^+$ and $\mu^-$ must have the same lifetime. Let us now consider some examples of non-metricity tensors that are common in the literature and analyze their contributions to the time dilation in this experiment.

%%%%%%%%%%%%%%%%%%%%%%%%%%%%%%%%%%%%%%%%%%%%%%%%%%%%%%%%%%%%%%%%%%%%%%%%%%%%%%%%%%%%%%%%%%%%%%%%%%%%%%%%%%%%%%%%%%%%%%%%%%%%%%%%%%%%%%%%%%%%%%%%%%%%%%%%%%%%%%%%%%%%%%%%%%%%%%%%%%%%%%%%%%%%%%%%%%%%%%%%%%%%%%%%

\subsection{Constant non-metricity}

As a first example, let us consider the case of a constant, universal non-metricity tensor. In 4 space-time dimensions, the non-metricity tensor can be decomposed in its Lorentz-irreducible parts as done in (\ref{NMdecomposed}) and (\ref{decompNM}). A decomposition that was also used to set experimental constraints on the irreducible components of the non-metricity tensor assuming a constant non-metricity background around Earth's surface \cite{Foster:2016uui}. Analyzing the expression (\ref{GeneralProperTimeEq}), we see that the second clock effect is sensitive to the contraction of the non-metricity tensor with the four-velocity of the clock in space-time. Since the contraction of all the indices of the tensor $M_{\mu\alpha\beta}$ with those of the four-velocity is null, it does not contribute to the effect. The terms proportional to $Q_{1\mu}$ and $Q_{2\mu}$ produce a contribution of the Weyl type, i.e., proportional to $\int \Sigma(\gamma')ds$, where $\boldsymbol{\Sigma}=(Q_{1\mu}+2Q_{2\mu})dx^{\mu}$. The symmetric tensor $S_{\mu\alpha\beta}$ furnishes an extra contribution to the effect. We can compute this contribution as follows. Consider the use of cylindrical coordinates and the four-velocity of a muon ($\mu^{\pm}$) in a circular trajectory with constant angular velocity and parametrized by the coordinate time $t$
\beq\label{4-vel}
\gamma'=\frac{d\gamma}{dt}=(c,0,\pm\dot{\theta},0)=(c,0,\pm\rho_0^{-1}\, v_0,0),
\eeq
where, as before, $\rho_0$ is the radius of the trajectory and $v_0$ is the modulus of its spatial velocity. Using these definitions in the proper time (\ref{GeneralProperTimeEq}) with the initial conditions
\beq
t_{0} =0\ \ , \ \ \left(\frac{d\tau/dt}{\sqrt{g(\gamma'(t)\gamma'(t))}}\right)_{t=t_{0}}=c^{-1},\label{init_cond}
\eeq
we can expand the exponential factor inside the integral in (\ref{GeneralProperTimeEq}) to find
\begin{align}
&\tau(t)=c^{-1}\int_0^t\sqrt{g(\gamma',\gamma')}du\\
&-\frac{c^{-1}}{2}\int_0^t\left[\int_0^{u}\frac{Q(\gamma',\gamma',\gamma')}{g(\gamma',\gamma',\gamma')}ds\right]\sqrt{g(\gamma',\gamma',\gamma')}du.\nonumber
\end{align}
\par
Now, we can decompose the non-metricity tensor in its irreducible parts, leading to
\begin{align}
Q(\gamma',\gamma',\gamma')=c^2\frac{\gamma^{-2}}{6}\Sigma_0V^0+c^2\frac{\gamma^{-2}}{6}\Sigma_2V^2+S_{000}(V^0)^3\nonumber\\
+S_{222}(V^2)^3+3S_{002}(V^0)^2V^2+3S_{022}V^0(V^2)^2,
\end{align}
where we define the Lorentz factor $\gamma^{-2}=c^{-2}\, g(\gamma'\gamma')=1-(v_0)^2/c^2$ and $V^{\mu}$ as the components of the four-velocity $\gamma'$ (\ref{4-vel}).
\par
With these definitions, a straightforward calculation leads to
\begin{align}
\tau(t)=\gamma^{-1}\, t-\frac{t^2}{4}\gamma^{-1}\left(\frac{c}{6}\Sigma_0+c\gamma^2S_{000}\right)-\frac{3}{4}t^2\frac{\gamma}{c}\rho_0^{-2}(v_0)^2S_{022}\nonumber\\
\mp \frac{t^2}{4}\left[\frac{\gamma^{-1}}{6}\rho_0^{-1}v_0\Sigma_2+\gamma\rho_0^{-3}\frac{(v_0)^3}{c^2}\, S_{222}+3 \gamma\rho_0^{-1}v_0\, S_{002}\right].
\end{align}

As before, in order to find the dilated lifetime we need to find $t$ as a function of $\tau$. Since $\gamma\, \tau$ is the dilated lifetime as predicted by special relativity, we can find a correction and a split in the dilated lifetimes of muons as
\begin{align}\label{dil-time3}
&t^{\text{Q}}(\mu^{\pm})=t^{\text{SR}}(\mu^{\pm})+c\frac{(t^{\text{SR}})^2}{4}\left(\frac{1}{6}\Sigma_0+\gamma^2S_{000}\right)\\
&+\frac{3}{4}(t^{\text{SR}})^2c^{-1}\rho_0^{-2}\gamma^2(v_0)^2S_{022}\nonumber\\
&\pm\frac{(t^{\text{SR}})^2}{4}\rho_0^{-1}v_0\left[\frac{1}{6}\Sigma_2+\gamma^2\rho_0^{-2}\frac{(v_0)^2}{c^2}\, S_{222}+3 \gamma^2 S_{002}\right].\nonumber
\end{align}
where $t^{\text{SR}}\doteq \gamma \tau$ is the dilated lifetime predicted by special relativity.\\

We can go further by noticing that since we are in cylindrical coordinates, the covectors that decomposed $Q_{\alpha\mu\nu}$ do not have the same units. So, we should transform to Cartesian coordinates in order to compare the different components of the non-metricity tensor. Also, it is reasonable to assume that, in Cartesian coordinates, the components of the non-metricity tensor are all of the same order of magnitude, just like is the case of the metric components. Since we are just interested in a rough constraint on the non-metricity tensor, we notice that the Jacobian matrix that shall transform the coordinates from cylindrical to Cartesian are of the order of the radius of the particle trajectory $\rho_0$. Therefore, in order to give such constraint we can simply correct the dimensions of the non-metricity components using powers of $\rho_0$. In fact, if $L$ represents the length dimension of a given quantity, a straightforward dimensional analysis reveals the following dimensionality for the components of $\boldsymbol{Q}$-decomposition
\begin{align}
&[\Sigma_0]=[S_{000}]=L^{-1} \ \ , \ \ [\Sigma_2]=[S_{002}]=1\, ,\\
&[S_{022}]=L \ \ , \ \ [S_{222}]=L^2.\nonumber
\end{align}

Thus, if we define new variables with the same dimensions as the non-metricity tensor, i.e., $L^{-1}$ (recall the definition of the non-metricity tensor $\nabla_{\alpha}g_{\mu\nu}=Q_{\alpha\mu\nu}$), we shall be able to constraint $\boldsymbol{Q}$. As we will see, this approach will be justifiable by comparison with previous cases \cite{Lobo:2018zrz}.
\par
Let us define
\begin{align}
&\tilde{\Sigma}_0=\Sigma_0 \ \ , \ \ \tilde{\Sigma}_2=\rho_0^{-1}\Sigma_2 \ \ , \ \ \tilde{S}_{000}=S_{000},\\
&\tilde{S}_{002}=\rho_0^{-1}S_{002} \ \ , \ \ \tilde{S}_{022}=\rho_0^{-2}S_{022} \ \ , \ \ \tilde{S}_{222}=\rho_0^{-3}S_{222}.\nonumber
\end{align}

As expected, each angular index ``$2$'' carries a power of $\rho_0^{-1}$ for these covectors. With these definitions, we should have these new variables of the same order of magnitude. For simplicity, let us assume that
\beq
\tilde{\Sigma}\sim \tilde{S}\sim Q,
\eeq
where $Q$ is a constant parameter with dimensions $L^{-1}$. This way, from the data of the CERN experiment also used in \cite{Lobo:2018zrz}, and the difference between the theoretical prediction of special relativity and the experimental uncertainty\footnote{It is worth stressing that such difference lies within the statistical error and can also be described by systematic effects.} $\Delta\tau(\mu^{\pm})\approx {\cal O}(10^{-2})\mu \text{s}$, and assuming that muons can describe a generalized clock, we could set a constraint on the absolute value of the parameter $Q$. 
\par
From
\begin{widetext}
\beq\label{difference}
|t^Q-t^{\text{SR}}|\sim Q\frac{(t^{\text{SR}})^2}{24\, c^2}\left|c^3+6\, c^3\gamma^2+18\, c\, \gamma^2(v_0)^2 \pm\, \left[6\gamma^2(v_0)^3 + c^2v_0 + 18c^2\gamma^2v_0\right]\right|\sim 10^{-2}\mu \text{s},
\eeq
\end{widetext}
we find an upper bound on the intensity of a universal non-metricity tensor (where we use $v_0=c\sqrt{1-\gamma^{-2}}$):
\beq
|Q|\lessapprox {\cal O}(10^{-14})\text{cm}^{-1}.
\eeq

The constraint found in \cite{Lobo:2018zrz} can be derived as special case of the above analysis if one assumes a relation between the Weyl component of the non-metricity tensor and the electromagnetic field. In fact, the contribution from Weyl geometry from that paper is given exclusively by the $c^2 v_0$ term of (\ref{difference}), which gives $|Q|\lessapprox {\cal O}(10^{-9})\text{cm}^{-1}$. However, $Q$ is given by $3|\lambda| \rho_0 B$, where $\lambda$ is the characteristic dimension-full constant of the model relating the electromagnetic field to the non-metricity, and the length and magnetic field scales of the muon experiment are respectively $\rho_0=700\, \text{cm}$ and $B=1.4\times 10^4 \text{G}$. Implying that $|\lambda|\approx 10^{-16}\, \text{G}^{-1}\, \text{cm}^{-2}$, which is the result found in \cite{Lobo:2018zrz}.\\

The effect of the general non-metricity tensor is governed by the $c^3 \gamma^2$ term of (\ref{difference}), which amplifies the second clock effect in comparison to the Weyl case, furnishing a much narrower constraint. In fact, if we divide this upper bound by the length and magnetic field scales of the muon experiment, we would find $10^{-21}\, \text{G}^{-1}\, \text{cm}^{-2}$. Improvements of this bound shall come from the Muon $g-2$ Fermilab experiment \cite{Venanzoni:2014ixa} that has been similarly designed and is currently investigating the mysterious anomalous magnetic moment of the muon \cite{Teubner-2018}, and should also publish the most precise data on the dilation of muons lifetime.

%%%%%%%%%%%%%%%%%%%%%%%%%%%%%%%%%%%%%%%%%%%%%%%%%%%%%%%%%%%%%%%%%%%%%%%%%%%%%%%%%%%%%%%%%%%%%%%%%%%%%%%%%%%%%%%%%%%%%%%%%%%%%%%%%%%%%%%%%%%%%%%%%%%%%%%%%%%%%%%%%%%%%%%%%%%%%%%%%%%%%%%%%%%%%%%%%%%%%%%%%%%%%%%%

\subsection{Ricci-based gravity}\label{sec:RicciBased}
In Ref.\cite{Avalos:2016unj}, it is found that a Weyl integrable space-time (WIST) is the most general Weyl space in which a Perlick clock does not measure a second clock effect. Since we have generalized this result for spaces with arbitrary non-metricity, it is pertinent to ask whether generalized clocks would measure a second clock effect in specific theories of modified gravity, or on the contrary, these theories give rise to space-times which are free of it. In this regard, let us notice that generalized clocks would measure a second clock effect in any theory of gravity which does not have a non-metricity tensor of the form $\boldsymbol{Q}=\dif\phi\otimes \boldsymbol{g}$. Among the most popular classes of theories, let us point out that the extended teleparallel theories (see e.g. \cite{Aldrovandi:2013wha}), which have vanishing curvature and non-metricity, are free of second clock effect. On the other hand, generalized clocks in the so-called symmetric teleparallel theories will in general measure a second-clock effect. The case of Ricci-Based gravity theories (RBGs), with a Lagrangian of the form $\mathcal{L}\lr{g_{\mu\nu}, R_{(\mu\nu)}(\Gamma)}$ should be further analyzed, as it is not so clear. RBGs can be defined as the most general class of projective invariant theories in which the action is an arbitrary analytic function of the metric and the Ricci tensor \cite{Afonso:2017bxr,BeltranJimenez:2017doy}. The projective symmetry is achieved by including only the symmetric part of the Ricci tensor in the action, and it ensures the absence of ghosts in these theories \cite{BeltranJimenez:2019acz,BeltranJimenezLargo}. A common characteristic of RBGs is the fact that they all possess an Einstein frame representation, and that the connection can be algebraically solved as the Levi-Civita connection of an auxiliary metric $q_{\mu\nu}$ which is related to the space-time metric by $g_{\mu\nu}= q_{\alpha\nu}({\Omega^{-1}})^\alpha{}_\mu$. The matrix $({\Omega^{-1}})^\alpha_{}{\beta}$, usually named {\it deformation matrix}, is generally determined by the relation $\sqrt{-|q|}q^{\mu\nu}=\sqrt{-|g|}\partial \mathcal{L}_G/\partial R_{\mu\nu}$, being $\mathcal{L}_G$ the gravitational Lagrangian. In general, this allows to build an equivalence between these theories and GR with a modified matter sector \cite{Afonso:2018bpv,Afonso:2018hyj,Afonso:2018mxn,Delhom:2019zrb}. In fact, this matrix has the property that it can be expanded as an analytic function of the stress-energy tensor, taking the form

\begin{align}\label{ExpansionOmega}
&(\Omega^{-1})_{\mu}{}^\al=\\
&\delta_\mu{}^\al+\sum^\infty_{n=1}\frac{1}{\Lambda^{4n}}\lrsq{\alpha_n T^n \delta_{\mu}{}^\al+\sum_{i=1}^n\beta^{(i)}_n f^{(n)}_{(n-i)}(T_{\mu}{}^\al)^{(i)}}.\nonumber
\end{align}

Here $(\al_n,\be^{(i)}_n)$ are the expansion coefficients that depend on the gravitational action, $(T_{\mu}{}^\al)^{(i)}\doteq T^\mu{}_{\nu_1}T^{\nu_1}{}_{\nu_2}...T^{\nu_i}{}_\al$ and $f^{(n)}_{(n-i)}$ are scalar functions built with $(n-i)$ powers of the stress-energy tensor.  It is worth noting that, given that in these theories $\nabla \boldsymbol{q}=0$, the non-metricity tensor can be written as
\beq\label{ExpansionNM}
Q_{\mu\al\be}=\lr{\nabla_\mu\Omega^{-1}_{\al}{}^\sigma}q_{\sigma\be}=\lr{\nabla_\mu\Omega^{-1}_{\al}{}^\sigma}\Omega_\sigma{}^\rho g_{\rho\be}.
\eeq

It is an interesting question to ask what is the subset of RBG theories that are free of the second clock effect. From the discussion in section (\ref{sec:SecondClock}) this question is equivalent to the question of what RBG theories yield Weyl integrable space-times. Using the relation between $q_{\mu\nu}$ and $g_{\mu\nu}$ and the form of the non-metricity tensor given in \eqref{ExpansionNM}, a necessary and sufficient condition for a specific RBG to be safe from the second-clock effect is the existence of a scalar function of $\phi(T)$ that satisfies
\beq\label{RBGsafeCondition}
\nabla_\mu\Omega^{-1}_{\al}{}^\sigma= (\dif\phi)_\mu\Omega^{-1}_{\al}{}^\sigma.
\eeq
Taking the trace in the two last indices of \eqref{RBGsafeCondition} we have that, if such function exists, it must satisfy
\beq\label{1-formForm}
\dif\phi=\text{d}\lr{\log \Omega^{-1}_\al{}^\al}.
\eeq
Without adding extra structure in our theory, or setting some ad-hoc constraints in the form of the stress energy tensor (which might be physically non-viable), the most general solution to the condition \eqref{RBGsafeCondition} is 
\beq\label{ConditionOmega}
\Omega^{-1}_{\al}{}^\sigma=e^{\phi+C}\delta_\al{}^\be,
\eeq
where $C$ is a constant. Therefore, we arrive to the conclusion that, in order for an RBG theory to avoid the second clock effect,  $g_{\mu\nu}$ and $q_{\mu\nu}$ must be conformally related,\footnote{From the point of view of Weyl transformations, $g_{\mu\nu}$ and $q_{\mu\nu}$  describe the same Weyl space-time since they are related by a Weyl transformation, thus being in the same conformal class.} which is true iff all $\beta^{(i)}_n$ coefficients in the expansion \eqref{ExpansionOmega} vanish.  As is well established, the subset of RBG theories in which $q_{\mu\nu}$ and $g_{\mu\nu}$ are conformally related is composed of all the metric-affine (or Palatini) $f(\mathcal{R})$ theories \cite{Olmo:2011uz}. To see this, notice that in general $\Omega^\alpha{}_\beta$ is defined to be proportional to $g_{\gamma\beta}\partial \mathcal{L}_G/\partial R_{\al\gamma}$, and $f(\mathcal{R})$ are the only RBG theories that satisfy $\partial \mathcal{L}_G/\partial R_{\al\gamma}\propto g^{\al\gamma}$, thus having a conformal deformation matrix. Therefore, among RBG theories, generalized clocks will in general measure a second clock effect, with the exception of $f(\mathcal{R})$ theories.\\

As an example, we can do the same analysis done in the last section for space-times with a non-metricity as the one generated in RBGs. This is, on Earth's surface, up to $\mathcal{O}(\Lambda_Q^{-8})$ and neglecting Newtonian and post-Newtonian corrections, the non-metricity tensor of RBG theories is given by
\beq
Q_{\al\mu\nu}=\frac{1}{\Lambda_Q^4}\lr{\al(\partial_\alpha T)\eta_{\mu\nu}+\be\partial_\al T_{\mu\nu}}.
\eeq

For the muon experiment, let us consider the stress energy tensor of the constant magnetic field pointing in the $z$ direction. This means $T=0$ and $T_{\mu\nu}=B^2/8\pi\, \text{diag}(1,-1,1,1)$ in the laboratory rest frame (where we are using the definition of \cite{Jackson:1998nia}, chapter 11). Therefore, given that $B^2$ is covariantly constant to this order, we have no non-metricity and no second-clock effect at this order.

%%%%%%%%%%%%%%%%%%%%%%%%%%%%%%%%%%%%%%%%%%%%%%%%%%%%%%%%%%%%%%%%%%%%%%%%%%%%%%%%%%%%%%%%%%%%%%%%%%%%%%%%%%%%%%%%%%%%%%%%%%%%%%%%%%%%%%%%%%%%%%%%%%%%%%%%%%%%%%%%%%%%%%%%%%%%%%%%%%%%%%%%%%%%%%%%%%%%%%%%%%%%%%%%

\section{Concluding remarks}\label{sec:Conc}

In this paper we defined the notion of generalized proper time for spaces with arbitrary non-metricity.  This can be done in a purely geometrical way, requiring only primitive geometrical objects like curves, a metric, and a connection. This definition naturally incorporates scale invariance and generalizes Perlick's definition for spaces with non-metricity more general than Weyl's. We also discussed the relation between this generalized proper time and the operational notion of time given by Ehlers, Pirani and Schild, and discussed how a second clock effect would arise for a generalized clock in presence of arbitrary non-metricity. This effect would be measurable provided that we find matter fields that allow us to construct a generalized clock, and under that assumption, we set bounds on several components of the non-metricity tensor from muon lifetime experiments at CERN. 
There is the question, however, of whether one can actually construct a generalized clock with the ingredients found in the universe, or whether it is just a mere idealization. To that end it is useful to define the notion of {\it physical clock} as a clock that measures time as experienced by physical observers ({\it i.e.} particles). By means of a generalized M\"arkze-Wheeler construction, one could in principle build clocks by using light rays and affine geodesics, instead of Riemannian ones, as done by M\"arkze and Wheeler in \cite{MarkzeWheeler,Desloge1989-DESATD}, and then study whether this construction bears any relation with generalized proper time. Notice that if it is to be related with generalized proper time, scale invariance should play a central role on this generalized M\"arkze-Wheeler construction. Nonetheless, even if the construction of generalized clocks is possible in this way, that does not guarantee that it is possible to find such clocks in nature. Particularly, one would need to find massive particles that follow geodesics of an affine connection with nontrivial non-metricity, thus implying a violation of the Einstein equivalence principle, which is strongly constrained experimentally \cite{Will:2014kxa}.  
 General relativity is constructed in such a way  that free test particles follow Riemannian geodesics due to the conservation of the energy-momentum tensor (which is also linked to the Bianchi identities) \cite{Geroch:1975uq}. This conservation law can also be deduced from the theory's invariance under diffeomorphisms (for instance, see section 4.1.8 of \cite{poisson2004relativist}) and depends on the definition of the energy-momentum tensor and in the way the matter Lagrangian is constructed, i.e., on the coupling between matter and geometry. Since, even in theories with non-trivial non-metricity, particles that are minimally coupled to the geometry \cite{FuturePaperMCP} follow Riemannian geodesics, they are not well suited as generalized clocks. Particularly, unless muons are seen to couple non-trivially to the affine structure, they will not be measuring generalized proper time. However, some proposals for different couplings have arisen recently. For instance, the case of integrable Weyl space-time was addressed in \cite{Romero:2012hs}, in which a coupling that obeys the gauge invariance of the geometry is proposed, and it turns out to be equivalent to the usual minimal coupling of general relativity in a Riemannian frame, thus implying that free particles follow Weyl geodesics in this theory. 
This issue was also addressed in \cite{deCesare:2016mml} for non-integrable Weyl geometry, and the authors concluded that free particles should follow Riemannian geodesics. In the context of $f(Q)$-gravity, it has been recently proposed  \cite{Harko:2018gxr} a coupling with matter that deserves an immediate analysis due to the appearance of an extra force in the geodesic equation (with the Levi-Civita connection). And more recently, the case of a non-minimal coupling between matter and geometry in manifolds endowed with a non-metric connection has gained growing attention \cite{Gomes:2018sbf,SanomiyaProceedings,Sanomiya:2020svg}. Thus, the issue of whether generalized proper time can be regarded as physical might depend on the particular model, since the possibility of constructing a generalized clock within every model depends on its particular geometry-matter coupling. Therefore further work in analysing different models is needed to see whether generalized clocks can (or cannot) be built naturally in any of them.\\

\section*{Acknowledgments}
A. D. is supported by a PhD contract of the program FPU 2015 (Spanish Ministry of Economy and Competitiveness) with reference FPU15/05406. G. J. O. is funded by the Ramon y Cajal contract RYC-2013-13019 (Spain).  This work is supported by the Spanish project FIS2017-84440-C2-1-P (MINECO/FEDER, EU), the project SEJI/2017/042 (Generalitat Valenciana), the project i-LINK1215 (A.E. CSIC), the Consolider Program CPANPHY-1205388, and the Severo Ochoa grant SEV-2014-0398 (Spain). I. P. L. and C. R. thank Conselho Nacional de Desenvolvimento Cient\'ifico e Tecnol\'ogico (CNPq) and Coordena\c c\~ao de Aperfei\c coamento de Pessoal de N\'ivel Superior (CAPES) - Finance Code 001, from Brazil, for financial support.

%%%%%%%%%%%%%%%%%%%%%%%%%%%%%%%%%%%%%%%%%%%%%%%%%%%%%%%%%%%%%%%%%%%%%%%%%%%%%%%%%%%%%%%%%%%%%%%%%%%%%%%%%%%%%%%%%%%%%%%%%%%%%%%%%%%%%%%%%%%%%%%%%%%%%%%%%%%%%%%%%%%%%%%%%%%%%%%%%%%%%%%%%%%%%%%%%%%%%%%%%%%%%%%%

\bibliography{Bibliography}
\bibliographystyle{JHEPmodplain}

\end{document}